\documentclass[aps,prd,preprint,superscriptaddress,tightenlines,nofootinbib]{revtex4}

\usepackage{graphicx}% Include figure files
\usepackage{dcolumn}% Align table columns on decimal point
\usepackage{bm}% bold math
\usepackage{epsfig}

\begin{document}

\newcommand{\ee}{e$^+$e$^-$}
\newcommand{\ff}{f$_{2}$(1525)}
\newcommand{\bb}{$b \overline{b}$}
\newcommand{\cc}{$c \overline{c}$}
\newcommand{\sbs}{$s \overline{s}$}
\newcommand{\uu}{$u \overline{u}$}
\newcommand{\dd}{$d \overline{d}$}
\newcommand{\qq}{$q \overline{q}$}
\newcommand{\suo}{\rm{\mbox{$\epsilon_{b}$}}}
\newcommand{\loro}{\rm{\mbox{$\epsilon_{c}$}}}
\newcommand{\kos}{\ifmmode \mathrm{K^{0}_{S}} \else K$^{0}_{\mathrm S} $ \fi}
\newcommand{\kol}{\ifmmode \mathrm{K^{0}_{L}} \else K$^{0}_{\mathrm L} $ \fi}
\newcommand{\ko}{\ifmmode {\mathrm K^{0}} \else K$^{0} $ \fi}

\def\tpc{three-particle correlation}
\def\twopc{two-particle correlation}
\def\ksks{K$^0_S$K$^0_S$}
\def\ee{e$^+$e$^-$}
\def\ff{f$_{2}$(1525)}

\preprint{CLNS 06/1977} % For paper draft CBX YY-NN -> Draft YY-NNA
\preprint{CLEO 06-19}         % for CLNS notes

\title{Confirmation of the $Y(4260)$ Resonance Production in ISR}

\author{Q.~He}
\author{J.~Insler}
\author{H.~Muramatsu}
\author{C.~S.~Park}
\author{E.~H.~Thorndike}
\author{F.~Yang}
\affiliation{University of Rochester, Rochester, New York 14627}
\author{T.~E.~Coan}
\author{Y.~S.~Gao}
\affiliation{Southern Methodist University, Dallas, Texas 75275}
\author{M.~Artuso}
\author{S.~Blusk}
\author{J.~Butt}
\author{J.~Li}
\author{N.~Menaa}
\author{R.~Mountain}
\author{S.~Nisar}
\author{K.~Randrianarivony}
\author{R.~Sia}
\author{T.~Skwarnicki}
\author{S.~Stone}
\author{J.~C.~Wang}
\author{K.~Zhang}
\affiliation{Syracuse University, Syracuse, New York 13244}
% ...
\collaboration{CLEO Collaboration} 
\noaffiliation

\date{October 25, 2006}

\begin{abstract} 
Using 13.3 fb$^{-1}$ of $e^+e^-$ collision data taken in the $\Upsilon(1S-4S)$ region with the CLEO III detector at the CESR collider, a search has been made for the new resonance $Y(4260)$ recently reported by the \textsc{BaBar} Collaboration. The production of $Y(4260)$ in initial state radiation (ISR), and its decay into $\pi^+\pi^-J/\psi$, are confirmed. A good quality fit to our data is obtained with a single resonance. We determine $M(Y(4260))=(4284^{+17}_{-16}(\mathrm{stat})\pm4(\mathrm{syst}))~\mathrm{MeV/{\it c}^{2}}$, $\Gamma(Y(4260))=(73^{+39}_{-25}(\mathrm{stat})\pm5(\mathrm{syst}))~\mathrm{MeV/{\it c}^{2}}$, and $\Gamma_{ee}(Y(4260))\times\mathcal{B}(Y(4260)\to\pi^+\pi^-J/\psi)=(8.9^{+3.9}_{-3.1}(\mathrm{stat})\pm1.8(\mathrm{syst}))$ eV/${\it c}^{2}$.  
\end{abstract}

\maketitle

The \textsc{BaBar} Collaboration \cite{ybabar} has reported the observation of a broad structure in the $\pi^+\pi^-J/\psi$ mass spectrum around 4.26 GeV/${\it c}^{2}$ in initial state  radiation (ISR) events, $e^+e^-\to\gamma_{\mathrm{ISR}}\pi^+\pi^-J/\psi$.  Data with a total luminosity of 233 $\mathrm{fb}^{-1}$ were taken at and near the peak of the $\Upsilon(4S)$ resonance.  The enhancement, labeled $Y(4260)$, was fitted as a single resonance with a signal/background ratio of approximately 2/1, number of events $N(Y)=125\pm13$, mass  $M(Y)=(4259\pm8(\mathrm{stat})^{+2}_{-6}(\mathrm{syst}))~\mathrm{MeV/{\it c}^{2}}$, width $\Gamma(Y)=(88\pm23(\mathrm{stat})^{+6}_{-4}(\mathrm{syst}))~\mathrm{MeV/{\it c}^{2}}$, and $\Gamma_{ee}(Y)\times\mathcal{B}(Y\to\pi^+\pi^-J/\psi)=(5.5\pm 1.0(\mathrm{stat})^{+0.8}_{-0.7}(\mathrm{syst}))~\mathrm{eV/{\it c}^{2}}$. The observation of $Y(4260)$ in ISR leads to the conclusion that it is a vector with $J^{PC}=1^{--}$, with additional evidence provided by the observation that the dipion mass spectrum peaks at high mass, consistent with a $S$--wave phase space model. Recently, \textsc{BaBar} has reported that no evidence for $Y(4260)$ excitation is found in the reaction $e^{+}e^{-} \to \gamma_{\mathrm{ISR}}p \bar{p}$, and obtained a ratio which is equivalent to a 90$\%$ confidence limit of $\Gamma_{ee}(Y)\times\mathcal{B}(Y\to p\bar{p})<0.72~\mathrm{eV/{\it c}^{2}}$ \cite{babar2}. In the reaction $B^{-} \to J/\psi \pi^{+}\pi^{-}K^{-}$ they report a 3.1$\sigma$ resonance signal and establish a 95$\%$ confidence limit of $\Gamma(B^{-} \to Y K^{-})\times\mathcal{B}(Y\to \pi^{+}\pi^{-}J/\psi)<255~\mathrm{eV/{\it c}^{2}}$ \cite{babar1}. None of the theoretical calculations of charmonium spectra predict a vector state near $4260~\mathrm{MeV/{\it c}^{2}}$, and the measurements of the parameter $R\equiv\sigma(\mathrm{hadrons})/\sigma(\mu^+\mu^-)$ actually show a dip in this region, which is quite well fitted with the known charmonium vectors \cite{seth}.  These facts make $Y(4260)$ an unexpected and provocative resonance, and many theoretical models have been proposed to explain its nature \cite{theory1,theory2}. This also makes it extremely important to make an independent experimental confirmation of the $Y(4260)$ resonance and its parameters. This paper reports on just such a confirmation.

In a recent measurement of direct $e^+e^-$ annihilation at $\sqrt{s}=4040$, 4160, and 4260 MeV/${\it c}^{2}$, CLEO \cite{ycleo} has confirmed the existence of an enhancement in the $\pi^+\pi^-J/\psi$ cross section at 4260 MeV/${\it c}^{2}$ by measuring a cross section consistent with the peak cross section reported by \textsc{BaBar}.  Similar enhancements were observed in $\pi^0\pi^0J/\psi$ and $K^+K^-J/\psi$.  However, because the near--peak $Y(4260)$ data were limited to a single energy point, the mass and width of $Y(4260)$ could not be determined.  In this paper, we present the results of the ISR production of $Y(4260)$ in 13.3 fb$^{-1}$ of $e^+e^-$ data taken in the $\Upsilon(1S-4S)$ region with the CLEO detector at CESR, the $e^+e^-$ collider at Cornell University.  Although our integrated luminosity is considerably less than that used by \textsc{BaBar}, the low background in our measurement enables us to make a statistically significant confirmation of $Y(4260)$, and an independent determination of its mass and width.

\begin{figure}[!tb]
\begin{center}
\includegraphics[width=3.in]{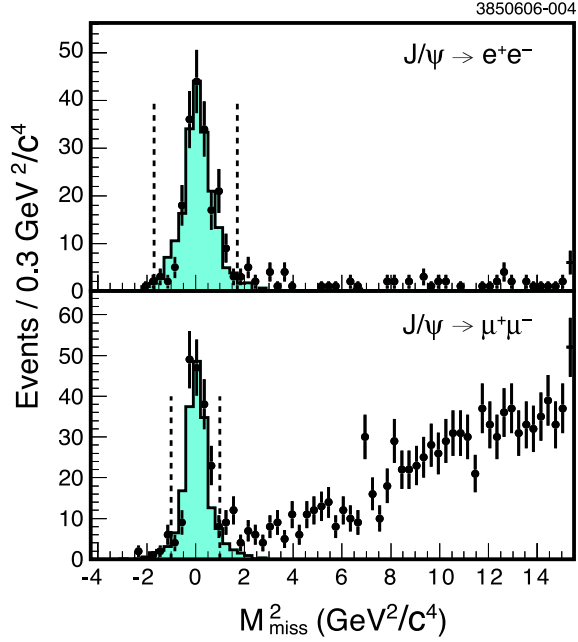}
\end{center}
\caption{Distributions of data events (points) for $M^2_{\mathrm{miss}}$, as defined in the text.  The peaks centered at $M^2_{\mathrm{miss}}=0$ and bounded by the dashed lines correspond to the missing ISR photons.  The shaded histograms show $\psi(2S)$ ISR signal Monte Carlo predictions normalized to the peak counts.}
\end{figure}

The data used in the present investigations consist of 1.40, 1.77, 1.60, and 8.57 fb$^{-1}$ taken at and in the vicinity of the $\Upsilon(1S)$, $\Upsilon(2S)$, $\Upsilon(3S)$, and $\Upsilon(4S)$ resonances, respectively.  The CLEO detector in its CLEO III \cite{cleodetector} configuration was used to detect final state particles.  The reaction studied was 
$$e^+e^-\to\gamma_{\mathrm{ISR}}\pi^+\pi^-J/\psi,~~J/\psi\to e^+e^-,~\mu^+\mu^-.$$
The four charged particles in the final state, $\pi^+\pi^-$ and $l^+l^-$, were detected. It was required that both electrons, or at least one muon be identified in the barrel or end cap regions of the detector. The energy $E$ deposited in the calorimeter, and the momentum $p$ measured by the tracking chamber, were used to form $E/p$.  It was required that for electrons $0.85<E/p<1.1$, and for muons $E/p<0.3$.  Electron candidates were required to have specific ionization energy loss, $dE/dx$, in the drift chamber consistent with the electron hypothesis within $3\sigma$. For muons a penetration depth of $\geq$3 hadronic interaction lengths in the muon chambers was required. Pions were identified using both the $dE/dx$ information from the drift chamber and the information from the Ring Imaging Cherenkov (RICH) detector.  The four charged particles were required to have net charge equal to zero. Further, very loose requirements were used, namely, total energy $<6.5$ GeV, total transverse momentum, $p_T<3.5$ GeV/{\it c}, and total energy of showers not matched with charged particles, $E_{\mathrm{neut}}<5$ GeV. Lepton pairs were accepted in the range $|M(l^+l^-)-M(J/\psi)|\le50$ MeV/${\it c}^{2}$.  The event selection criteria were optimized by using GEANT-based \cite{geant} $Y(4260)$ ISR signal Monte Carlo and $J/\psi$ side band yield as background, to maximize $S^2/(S+B)$, where $S$ and $B$ refer to the signal and background event counts, respectively.  The sideband was defined as 2.7 to 3.5 GeV/${\it c}^{2}$, excluding the region 2.9 to 3.2 GeV/${\it c}^{2}$.

The ISR events were identified by constructing the mass of the missing particle as
$$M^2_{\mathrm{miss}} = (2E_{\mathrm{beam}}-E(\pi^+\pi^- J/\psi))^2 - p(\pi^+\pi^-J/\psi)^2.$$
The resulting $M^2_{\mathrm{miss}}$ distributions are shown in Fig.~1. The peaks centered at $M^2_{\mathrm{miss}}=0$ correspond to the missing ISR photon and have little background under them.  Above the peak in the spectrum for $J/\psi\to\mu^+\mu^-$, the background increases rapidly because of misidentified pions. We consider events with $|M^2_{\mathrm{miss}}|<1.7~\mathrm{GeV^{2}/{\it c}^{4}}$ for $J/\psi\to e^+e^-$, and events with $|M^2_{\mathrm{miss}}|<1.0~\mathrm{GeV^{2}/{\it c}^{4}}$ for $J/\psi\to\mu^+\mu^-$, as ISR events.  In order to improve the mass resolution, these events were fitted with the lepton tracks constrained to a common vertex, and the dilepton mass constrained to the $J/\psi$ mass, $3097$ MeV/${\it c}^{2}$.  The distributions for the $\chi^2$ of the vertex and $J/\psi$ mass constraints, lepton and pion momenta, transverse momentum, and neutral energy for these selected events were found to be in good agreement with the corresponding distributions for the $\psi(2S)$ ISR signal Monte Carlo (more than 95\% of the ISR events come from $\psi(2S)$).  In Fig.~2 we show the $M(\pi^+\pi^-J/\psi)$ distributions for these events for $J/\psi\to e^+e^-$ and $\mu^+\mu^-$.  Both plots show clear enhancements around 4260 MeV/${\it c}^{2}$, with very little background.  This background is found to be in good agreement with that determined from the sidebands of $J/\psi$.

\begin{figure}[!tb]
\begin{center}
\includegraphics[width=3.in]{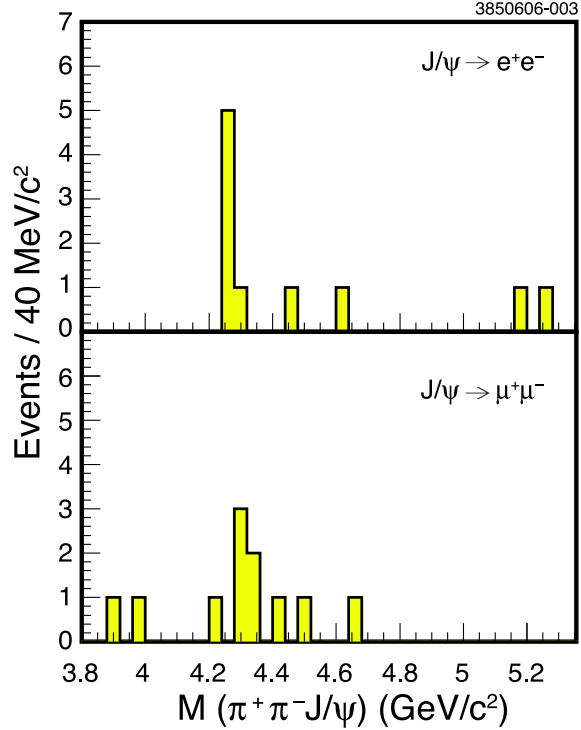}
\end{center}
\caption{Event distributions after kinematic fitting for the invariant mass $M(\pi^+\pi^-J/\psi)$, with $J/\psi\to e^+e^-$ and $\mu^+\mu^-$.  Evidence for the production of Y(4260) is present in both distributions.}
\end{figure}

The mass resolution was determined by fitting the ISR signal Monte Carlo distributions for $M(\pi^+\pi^-J/\psi)$ for $\psi(2S)$ and $Y(4260)$, assuming zero intrinsic width for both.  The instrumental resolution, full--width at half--maximum, was thus determined to be 8.2 MeV/${\it c}^{2}$ for $\psi(2S)$ and 10.6 MeV/${\it c}^{2}$ for $Y(4260)$. The data for $\psi(2S)$ and Y(4260) enhancements were each fitted with a constant background plus the peak shape determined by the Monte Carlo simulation convoluted with the single resonance relativistic Breit-Wigner function which includes the phase space factor.  The mass, width, peak area, and the background were kept free in the fits.  The fit for the $\psi(2S)$ peak returned $N(\psi(2S))=285\pm17$ counts and $M(\psi(2S))=(3685.70\pm0.24)$ MeV/${\it c}^{2}$, which compares favorably with the known mass, $M(\psi(2S))=(3686.093\pm0.034)$ MeV/${\it c}^{2}$ \cite{pdg}. The fit for the $Y(4260)$ enhancement is illustrated in Fig.~3.  It results in $M(Y(4260))=(4284^{+17}_{-16})$ MeV/${\it c}^{2}$ and $\Gamma(Y(4260))=(73^{+39}_{-25})$ MeV/${\it c}^{2}$.  The number of counts in the fitted resonance is $N(Y(4260))=13.6^{+4.7}_{-3.9}$.  The fit with a single Breit--Wigner has $\chi^2/d.o.f.=1.00$, a confidence level of 52\%, and a significance of $5.4\sigma$. The significance level for the resonance was obtained as $\sigma\equiv\sqrt{-2 \ln(L_0/L_{\mathrm{max}})}$, where $L_{max}$ is the maximum likelihood value for the fit with resonance, and $L_0$ is the likelihood value for the fit without resonance.   

\begin{figure}[!tb]
\begin{center}
\includegraphics[width=3.in]{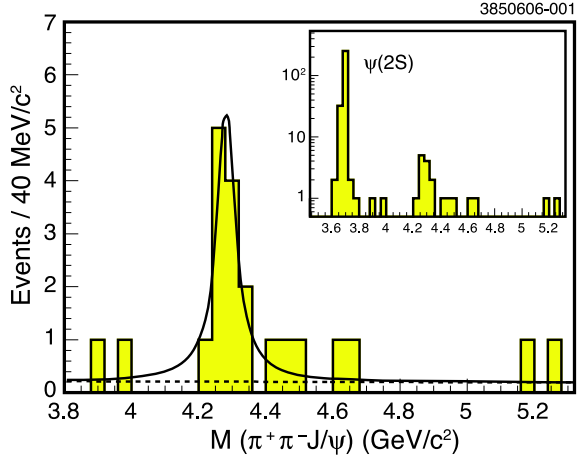}
\end{center}
\caption{The $M(\pi^+\pi^-J/\psi)$ distribution for the sum of $J/\psi\to e^+e^-$ and $J/\psi\to\mu^+\mu^-$.  The dotted line denotes the fitted background, and the solid curve denotes the fit with a single resonance.  The semilog plot in the insert illustrates the $\psi(2S)$ excitation in the extended mass region.}
\end{figure}

The systematic errors in our measurements of $M(Y(4260))$ and $\Gamma(Y(4260))$ are much smaller than the statistical errors.  An estimate of the systematic error in our mass measurement for $Y(4260)$ is provided by our result for $M(\psi(2S))$.  As mentioned earlier, our measured mass $M(\psi(2S))$ differs from its known value by only ($0.4\pm0.2$) MeV/${\it c}^{2}$.  Additional contributions to the systematic uncertainty in $M(Y(4260))$ and $\Gamma(Y(4260))$ were estimated by fitting the $Y(4260)$ peak in the data by varying by $\pm$50$\%$ the resolution width returned by our ISR Monte Carlo sample. We have also studied the effect of variations in event selection cuts ($M^2_{\mathrm{miss}}$, $p_T$, $E_{\mathrm{neut}}$, $\chi^2$ for vertex and $M(J/\psi)$ constraints, lepton identification criteria), and variations in peak shape and background parameterization.  The maximum variation found was $\pm$4 MeV/${\it c}^{2}$ in $M(Y(4260))$ and $\pm$5 MeV/${\it c}^{2}$ in $\Gamma(Y(4260))$, which we therefore assign as systematic uncertainties.

\begin{figure}[!tb]
\begin{center}
\includegraphics[width=3.in]{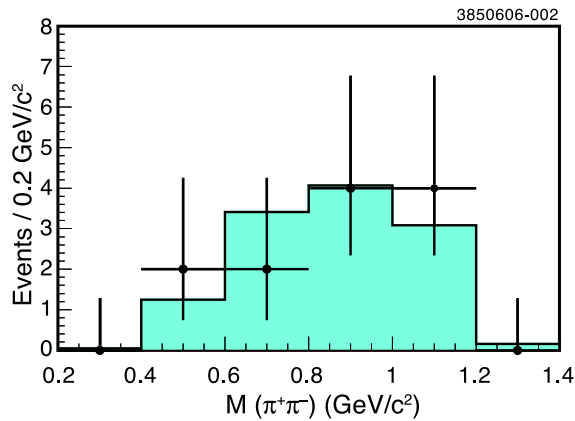}
\end{center}
\caption{Dipion mass distribution for data events (points).  The shaded histogram is the Monte Carlo prediction for $M(\pi^+\pi^-)$ based on a $S$--wave phase space model. Neither the data nor the Monte Carlo distributions have been corrected for the experimental efficiency.}
\end{figure}

Thus our final results for the mass and width are
\begin{displaymath}
M(Y(4260))=(4284^{+17}_{-16}\mathrm{(stat)}\pm4\mathrm{(syst)})~\mathrm{MeV/{\it c}^{2}},
\end{displaymath}
\begin{displaymath}
\Gamma(Y(4260))=(73^{+39}_{-25}\mathrm{(stat)}\pm5\mathrm{(syst)})~\mathrm{MeV/{\it c}^{2}}.
\end{displaymath}

The quantity $\Gamma_{ee}(Y)\times \mathcal{B}(Y \to \pi^+\pi^-J/\psi)$
is related to the observed counts $N(Y)$ as
$$\Gamma_{ee}(Y) \times \mathcal{B}(Y \to \pi^+\pi^-J/\psi)=
\frac{N(Y)/\mathcal{B}(J/\psi \to l^{+}l^{-})}
{\mathcal{L} \times \epsilon (Y) \times \sigma_{\mathrm{ISR}}(Y)},$$
where $\mathcal{L}$ is the integrated luminosity, $\epsilon(Y)$ is the  
efficiency determined by the ISR signal Monte Carlo, 
and $\sigma_{\mathrm{ISR}}(Y)$ is the value of the ISR cross section 
per keV/${\it c}^{2}$ of di-electron width calculated using the theoretical formalism of Kuraev and Fadin \cite{isrformal}.  

From our sample of $\psi(2S)$ ISR events, for which $\Gamma_{ee}(\psi(2S)) \times \mathcal{B} (\psi(2S) \to \pi^+\pi^-J/\psi)$ is known \cite{pdg}, we find that our ISR signal Monte Carlo efficiency $\epsilon(\psi(2S))$ is overestimated by a factor of 1.5. We attribute this to the fact that because of the extreme forward nature of the ISR events the geometrical acceptance of our detector ($|\cos \theta|<$0.93) allows us to accept only approximately 15$\%$ of the phase space for $\psi(2S)$, and we become extremely sensitive to the Monte Carlo modeling of the ISR production mechanism. For $Y(4260)$ events the geometrical acceptance of our detector is larger, approximately 30$\%$, and we therefore expect that the Monte Carlo determined efficiency for the $Y(4260)$ events is more accurate than that for the $\psi(2S)$ events. We estimate the appropriate efficiency correction for the $Y(4260)$ events to be between 1.0 and 1.5, and we assume it to be 1.25$\pm$0.25. This is the dominant systematic uncertainty in our measurement of $\Gamma_{ee}(Y) \times \mathcal{B}(Y \to \pi^+\pi^-J/\psi)$. Systematic uncertainties in luminosity, $\sigma_{\mathrm{ISR}}$, and $\mathcal{B}(J/\psi \to l^{+}l^{-})$ are much smaller. To further minimize the systematic uncertainty in our determination of $\Gamma_{ee}(Y) \times \mathcal{B}(Y \to \pi^+\pi^-J/\psi)$ we impose the additional requirement for its estimate that at least one lepton from the $J/\psi$ decay have $|\cos \theta|<$0.7. As a result, the data were refitted for the Y(4260) resonance.  The fit returned $N(Y)=8.1^{+3.6}_{-2.9}$ with a $\chi^2/d.o.f.=0.98$.  With $\sigma_{\mathrm{ISR}}(Y)=6.26~\mathrm{pb/keV}$, and $\epsilon(Y)=11.5\%/(1.25\pm0.25)$, our final result is
\begin{eqnarray*}
\Gamma_{ee}(Y) \times \mathcal{B} (Y \to \pi^+\pi^-J/\psi)\hspace{4.cm}\\
\hspace{1.cm}=(8.9^{+3.9}_{-3.1}\mathrm{(stat)}\pm1.8\mathrm{(syst)})~\mathrm{eV/{\it c}^{2}.}
\end{eqnarray*}
This is consistent with the \textsc{BaBar} result.

Fig.~4 shows that the dipion mass of the $Y(4260)$ events is enhanced at high mass, 
as predicted by Monte Carlo generated distribution based on a phase space model modified by rescattering of pions which are in a relative $S$ state \cite{cahn}.

In summary, we have confirmed \textsc{BaBar}'s observation of a new resonance, $Y(4260)$, 
produced in initial state radiation.  We obtain resonance parameters in agreement 
with theirs. A good quality fit to our data is obtained with a single resonance.
The observation of $Y(4260)$ production in ISR and the $S$--wave phase space distribution of the dipion mass confirm that $J^{PC}(Y(4260))=1^{--}$ \cite{ybabar,ycleo}. The various theoretical models proposed for $Y(4260)$ have been discussed in detail in reviews by Eichten, Lane and Quigg \cite{theory1}, and Swanson \cite{theory2}, but no clear understanding of the nature of $Y(4260)$ exists at present. A dedicated resonance scan and good measurements of various decays of $Y(4260)$ are needed to discriminate among the different theoretical models.

We gratefully acknowledge the effort of the CESR staff in providing us with 
excellent luminosity and running conditions. D.~Cronin-Hennessy and A.~Ryd thank 
the A.P.~Sloan Foundation. This work was supported by the National Science
Foundation, the U.S. Department of Energy,  and the Natural Sciences and 
Engineering Research Council of Canada.


\begin{thebibliography}{99}

\bibitem{ybabar} 
B.~Aubert \textit{et al.} (\textsc{BaBar} Collaboration), Phys. Rev. Lett. \textbf{95}, 142001 (2005).

\bibitem{babar2}
B.~Aubert \textit{et al.} (\textsc{BaBar} Collaboration), Phys. Rev. \textbf{D 73}, 012005 (2006).

\bibitem{babar1} 
B.~Aubert \textit{et al.} (\textsc{BaBar} Collaboration), Phys. Rev. \textbf{D 73}, 011101 (2006).

\bibitem{seth} 
K.~K.~Seth, Phys. Rev. \textbf{D 72}, 017501 (2005).

\bibitem{theory1}
E.~J.~Eichten, K.~Lane, and C.~Quigg, Phys. Rev. \textbf{D 73}, 014014 (2006).

\bibitem{theory2} E.~Swanson, Phys. Rept. \textbf{429}, 243 (2006).

\bibitem{ycleo} 
T.~E.~Coan \textit{et al.} (CLEO Collaboration), 
Phys. Rev. Lett. \textbf{96}, 162003 (2006).

\bibitem{cleodetector} 
Y.~Kubota \textit{et al.} (CLEO Collaboration), Nucl. Instrum. Meth. \textbf{A 320}, 66 (1992); M.~Artuso \textit{et al.}, Nucl. Instrum. Meth. \textbf{A 554}, 147 (2005); D.~Peterson \textit{et al.}, Nucl. Instrum. Meth. \textbf{A 478}, 142 (2002).

\bibitem{geant} 
R.~Brun \textit{et al.}, GEANT 3.15, CERN Report \#DD/EE/84--1, 1987.

\bibitem{pdg} 
W. -M.~Yao \textit{et al.} (Particle Data Group), J.Phys. \textbf{G 33}, 1 (2006).

\bibitem{isrformal} 
E.~A.~Kuraev and V.~S.~Fadin, Sov. J. Nucl. Phys. \textbf{41}, 466 (1985).


\bibitem{cahn} L.~S.~Brown and R.~N.~Cahn, Phys. Rev. Lett. \textbf{35}, 1 (1975).

\end{thebibliography}
\end{document}